\documentclass[aps,pra,reprint,amsmath,amssymb,superscriptaddress,nobibnotes, longbibliography]{revtex4-1}
\usepackage{graphicx,SIunits}
\usepackage{bm}     
\usepackage{hyperref} 
\usepackage{dsfont}    
\usepackage{color}

\usepackage{enumitem}

\usepackage{amsmath}
\usepackage{amsthm}
\usepackage{amssymb}
\usepackage{tikz}

\usetikzlibrary{decorations.pathreplacing}
\usepackage{braket}
\usepackage{dsfont}

\usepackage{color}
\usepackage{lineno}
\usepackage{etoolbox}

\usepackage{multirow}

\begin{document}

\preprint{APS/123-QED}

\title{Contextuality of quantum non-demolition measurement via state discrimination}

  \author{Min Namkung}%
      \affiliation{Center for Quantum Technology, Korea Institute of Science and Technology (KIST), Seoul 02792, Korea}

  \author{Ilhwan Kim}
          \affiliation{Center for Quantum Technology, Korea Institute of Science and Technology (KIST), Seoul 02792, Korea}
  
  \author{Hyang-Tag Lim}
      \email{hyangtag.lim@kist.re.kr}
      \affiliation{Center for Quantum Technology, Korea Institute of Science and Technology (KIST), Seoul 02792, Korea}
      \affiliation{Division of Quantum Information, KIST School, Korea University of Science and Technology, Seoul 02792, Korea}

\date{\today}

\begin{abstract}
Quantum non-demolition measurements facilitate various quantum technologies, including quantum communication. Notably, their operational structure can be replicated by a classical model--referred to as a noncontextual model--making it crucial to identify which features prevents such models from reproducing the corresponding quantum measurements. In this work, we theoretically demonstrate contextual features inherent in the structure of quantum non-demolition measurements. These features not only reveal the nonclassicality of unambiguous state discrimination, but also extend to sequential unambiguous discrimination and probabilistic quantum cloning, both of which involve post-measurement states. Moreover, our analysis extends to noisy scenarios, highlighting its potential relevance for practical implementations. We believe that our results broaden the scope of observing nonclassicality in quantum systems and ultimately contribute to the advancement of various quantum technologies.
\end{abstract}


\maketitle

\begin{figure*}[t]
\centerline{\includegraphics[width=0.8\linewidth]{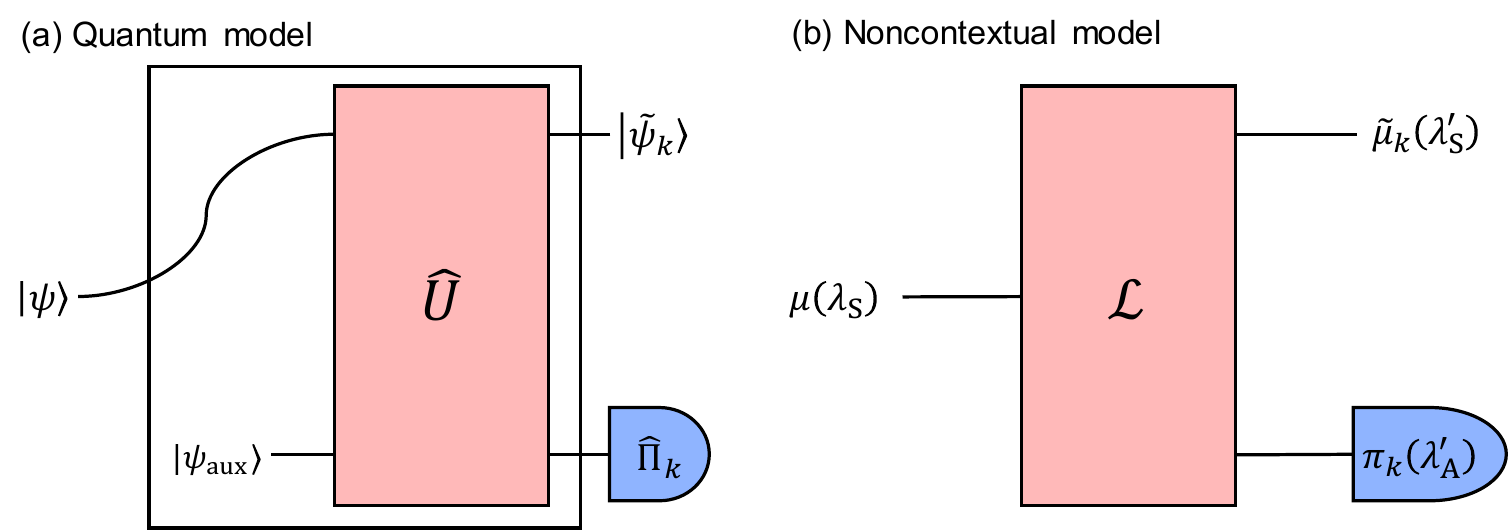}}
\caption{Structure of non-demolition measurement. (a) Structure in the quantum model, consisting of an auxiliary state $|\psi_{\rm aux}\rangle$, a unitary operator $\hat{U}$, and a direct measurement ${\hat{\Pi}_k}$. The post-measurement state is denoted by $|\widetilde{\psi}_k\rangle$. (b) Structure in a noncontextual ontological model, consisting of a transformation $\mathcal{L}$ and a direct measurement ${\pi_k(\lambda_{{\rm A}}')}$, yielding the post-measurement epistemic state $\widetilde{\mu}_k(\lambda_{{\rm S}}')$.}
\centering
\label{figure1}
\end{figure*}

\section{Introduction}
Rigorously identifying nonclassical characteristics of quantum systems holds great promise not only for studying the foundations of quantum mechanics~\cite{a.einstein,j.s.bell,j.f.clauser}, but also for developing various quantum technologies~\cite{a.k.ekert,h.j.briegel,h.lee}. Among such features, contextuality--initially proposed by Kochen and Specker~\cite{s.kochen,c.budroni} and later generalized by Spekkens~\cite{r.w.spekkens}--rules out hidden-variable models without requiring spacelike separation, as is typically assumed in nonlocality~\cite{n.brunner}. This distinctive form of nonclassicality has been shown to provide advantages in various quantum tasks such as quantum communication~\cite{s.gupta}, quantum computation~\cite{j.bowles}, and quantum sensing~\cite{j.jae}. In particular, contextuality leads to three notable consequences: it prevents the success probabilities of quantum state discrimination--such as minimum-error~\cite{c.w.helstrom,j.bae}, unambiguous~\cite{i.d.ivanovic,d.dieks,a.peres,g.jaeger}, maximal-confidence~\cite{s.croke}, and fixed-failure-rate discrimination~\cite{u.herzog}--from being reproduced by noncontextual models~\cite{d.schmid,s.mukherjee,k.flatt,j.shin2,k.flatt2,m.namkung_fr}, thereby enabling quantum random number generation~\cite{c.r.i.carceller}; it allows state-dependent cloning with fidelities beyond the noncontextual bound~\cite{m.lostaglio}]; and it reveals the nonclassicality of the Mach--Zehnder interferometric setup~\cite{r.wagner}, employed for quantum-enhanced phase estimation~\cite{h.lee}.

The aforementioned contextuality-related tasks are primarily formulated within a prepare-and-measure scenario between a sender and a receiver. Notably, quantum theory also allows for quantum non-demolition measurements~\cite{k.kraus}, which do not completely extract all the information from the prepared state~\cite{p.rapcan}. This enables various schemes such as sequential unambiguous state discrimination~\cite{j.a.bergou_seq} and probabilistic quantum cloning~\cite{l.m.duan_cl}, where post-measurement states play a crucial role. In sequential unambiguous state discrimination, which applies to quantum key distribution~\cite{c.h.bennett,m.namkung_qkd}, multiple receivers arranged in sequence are able to discriminate the sender's initial states without error. In probabilistic quantum cloning, it is possible to produce perfect copies with a certain probability. Moreover, quantum non-demolition measurements have been experimentally demonstrated in various platforms, such as optical systems~\cite{p.grangier} and spin systems~\cite{t.nakajima}. These observations suggest that contextual aspects of quantum non-demolition measurements may provide additional advantages for more sophisticated quantum information tasks that are experimentally accessible. However, one should be cautious: as in randomness certification, a noncontextual measurement formalism can deceptively mimic the quantum description~\cite{c.r.i.carceller,footnote}, despite fundamental differences.  This raises the following question: to what extent can quantum non-demolition measurements be reproduced by a noncontextual model, and where does this reproduction break down?

In this work, we theoretically demonstrate contextual features of quantum non-demolition measurements and systematically address the above question. We first formulate quantum non-demolition measurements within the noncontextual framework introduced by Spekkens~\cite{r.w.spekkens}. We then show that, although quantum non-demolition measurements can be equivalently described within both quantum and noncontextual models at the operational level, there exist regimes in which the noncontextual framework fails to reproduce the quantum predictions. We use this observation to establish contextual advantages in various protocols, including optimal unambiguous discrimination, sequential unambiguous discrimination, and probabilistic quantum cloning. Moreover, our formulation extends to noisy scenarios, highlighting its practical relevance, and clarifies the origin of contextuality enhancement in maximal-confidence discrimination~\cite{k.flatt,k.flatt2}. As these protocols are closely related to quantum sensing~\cite{m.hillery} and communication~\cite{g.cariolaro}, we expect that our results provide a new route toward enhancing the capabilities of a wide range of quantum technologies.

\section{Preliminaries}
We first briefly introduce the formulation of states, transformations, and measurements based on the noncontextual ontological model proposed in Refs.~\cite{r.w.spekkens,d.schmid}. In this framework, a state is described by a probability distribution over a hidden variable $\lambda \in \Omega$, denoted by $\mu(\lambda)$. Here, $\lambda$ and $\mu(\lambda)$ are referred to as the ontic state and the epistemic state, respectively. By definition, $\mu(\lambda)$ satisfies
\begin{align}
    \mu(\lambda) &\ge 0, \ \ \forall \lambda \in \Omega, \nonumber\\
    \int_{\Omega} d\lambda \, \mu(\lambda) &= 1.
\end{align}

A transformation from one state to another is described by a non-negative function $\mathcal{L}(\lambda', \lambda)$ satisfying
\begin{align}\label{LL}
    \int_{\Omega'} d\lambda' \, \mathcal{L}(\lambda', \lambda) = 1, \ \ \forall \lambda \in \Omega,
\end{align}
which induces the transformed epistemic state
\begin{align}
    \mu'(\lambda') = \int_{\Omega} d\lambda \, \mathcal{L}(\lambda', \lambda)\, \mu(\lambda).
\end{align}

A measurement $\mathcal{M}$ with discrete outcomes $k$ is described by response functions $\xi_k(\lambda)$ satisfying
\begin{align}\label{meas}
    \xi_k(\lambda) &\ge 0, \ \ \forall k, \ \forall \lambda \in \Omega, \nonumber\\
    \sum_k \xi_k(\lambda) &= 1, \ \ \forall \lambda \in \Omega.
\end{align}

Using this formalism, the probability of obtaining outcome $k$ when measuring the transformed epistemic state is given by
\begin{align}\label{prob} 
    p(k|\mathcal{M}, \mathcal{L}, \mu)
    &= \int_{\Omega'} d\lambda' \, \xi_k(\lambda') \mu'(\lambda') \nonumber\\
    &= \int_{\Omega' \times \Omega} d\lambda' d\lambda \, \xi_k(\lambda') \mathcal{L}(\lambda', \lambda) \mu(\lambda),
\end{align}
where ``$\times$'' denotes the Cartesian product. This expression shows that the framework provides an operational description of physical processes without invoking additional postulates.

\section{Results}
We now formulate non-demolition measurements within the noncontextual model in order to systematically determine the extent to which they can be reproduced by such a model. We focus in particular on discrimination tasks, including optimal unambiguous state discrimination~\cite{g.jaeger}, sequential unambiguous state discrimination~\cite{j.a.bergou_seq}, and probabilistic quantum cloning~\cite{l.m.duan_cl}.

\subsection{Replicating non-demolition measurement}
Remarkably, quantum non-demolition measurements take the following structure, as illustrated in Fig.~\ref{figure1}(a): preparation of an auxiliary system; interaction between the system to be measured and the auxiliary system; and a direct measurement on the auxiliary system to extract information about the initial state~\cite{m.a.neumark}. It is also worth noting that the auxiliary-state preparation and the interaction can be unified into a single quantum operation, represented by an isometry from a Hilbert space to an extended Hilbert space~\cite{p.-x.chen,y.sun,s.franke-arnold,m.agnew,s.goel,k.-m.hu}.

Under this structure, we reproduce the non-demolition measurement within the noncontextual model, as shown in Fig.~\ref{figure1}(b). Let us consider an initial epistemic state $\mu(\lambda_{\rm S})$ defined on a system $\rm S$. The state to be measured is transformed by a map $\mathcal{L}$ from the ontic space $\Omega_{\rm S}$ to the extended space $\Omega_{\rm S}' \times \Omega_{\rm A}'$, resulting in
\begin{align}\label{comp}
    \mu'(\lambda_{\rm S}', \lambda_{\rm A}') = \int_{\Omega_{\rm S}} d\lambda_{\rm S} \, \mathcal{L}(\lambda_{\rm S}', \lambda_{\rm A}', \lambda_{\rm S}) \mu(\lambda_{\rm S}).
\end{align}

The resulting epistemic state is locally measured by a direct measurement, described by a set of response functions $\Pi=\{\pi_k(\lambda_{\rm A}')\}$ on $\Omega_{\rm A}'$ satisfying
{\begin{align}\label{dir}
    \pi_k(\lambda_{\rm A}')\ge0,& \ \ \ \forall k, \ \forall\lambda_{\rm A}'\in\Omega_{\rm A}' \nonumber\\
    \sum_{k}\pi_{k}(\lambda_{\rm A}')=1,& \ \ \ \forall\lambda_{\rm A}'\in\Omega_{\rm A}'\nonumber\\
    \pi_j(\lambda_{\rm A}')\pi_k(\lambda_{\rm A}')=\pi_k(\lambda_{\rm A}')\delta_{jk},& \ \ \ \forall j, \ \forall k, \ \forall\lambda_{\rm A}'\in\Omega_{\rm A}',
\end{align}
with the Kronecker delta $\delta_{jk}$.} This implies that the support spaces between $\pi_k(\lambda_{\rm A}')$ and $\pi_{k'}(\lambda_{\rm A}')$ with $k\not=k'$ are orthogonal with each other. 

Substituting Eqs.~(\ref{comp}) and (\ref{dir}) into Eq.~(\ref{prob}), the probability of obtaining outcome $k$ is given by
\begin{align}\label{meas_prob}
    &p(k|\mathcal{M}_{\rm nd},\mu)=\int_{\Omega_{\rm S}'\times\Omega_{\rm A}'}d\lambda_{\rm S}'d\lambda_{\rm A}'\pi_k(\lambda_{\rm A}')\mu'(\lambda_{\rm S}',\lambda_{\rm A}')\nonumber\\
    &=\int\limits_{\substack{\Omega_{\rm S}'\times\Omega_{\rm A}'\\ \times\Omega_{\rm S}}}d\lambda_{\rm S}'d\lambda_{\rm A}'d\lambda_{\rm S}\pi_k(\lambda_{\rm A}')\mathcal{L}(\lambda_{\rm S}',\lambda_{\rm A}',\lambda_{\rm S})\mu(\lambda_{\rm S}),
\end{align}
where $\mathcal{M}_{\rm nd} = (\Pi, \mathcal{L})$ denotes the non-demolition measurement. This description is compatible with the general measurement formalism in Eq.~(\ref{meas}), as discussed in Appendix~A.1.

We further note that, since the isometry in quantum non-demolition measurements preserves the fidelity between quantum states {on a closed system}, the transformation $\mathcal{L}$ must preserve
\begin{align}\label{cp}
    \Big(\mu(\lambda_{\rm S}),\nu(\lambda_{\rm S})\Big)_{\Omega_{\rm S}}=\Big(\mu'(\lambda_{\rm S}',\lambda_{\rm A}'),\nu'(\lambda_{\rm S}',\lambda_{\rm A}')\Big)_{\Omega_{\rm S}'\times\Omega_{\rm A}'},
\end{align}
where $\big(\mu(\lambda), \nu(\lambda)\big)_{\Omega}$ denotes the confusability between $\mu(\lambda)$ and $\nu(\lambda)$ on $\Omega$, defined as~\cite{d.schmid}
\begin{align}\label{con_p}
    \big(\mu(\lambda), \nu(\lambda)\big)_{\Omega}
    &= \int_{\mathrm{supp}[\mu(\lambda)]} d\lambda \, \nu(\lambda) \nonumber\\
    &= \int_{\mathrm{supp}[\nu(\lambda)]} d\lambda \, \mu(\lambda),
\end{align}
with $\mathrm{supp}[\mu(\lambda)]$ denoting the support of $\mu(\lambda)$. The confusability quantifies the overlap between two epistemic states and corresponds to the overlap $|\langle \psi_\mu | \psi_\nu \rangle|^2$ between the associated quantum states~\cite{d.schmid}. {We note that the transformation $\mathcal{L}$ acts on a closed system, and thus any increase or decrease in confusability arising from information exchange with an external system is not allowed. Therefore, it is rational to consider that the confusability is preserved as described by Eq.~(\ref{cp}).}

\begin{figure}[t]
\centerline{\includegraphics[width=1\linewidth]{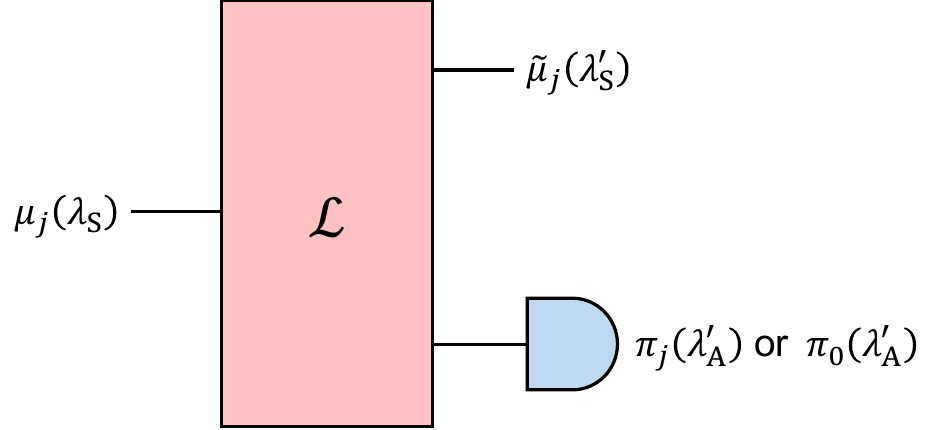}}
\caption{Non-demolition measurement reproduced within the noncontextual model for performing unambiguous discrimination between two epistemic states $\mu_1(\lambda_{\rm S})$ and $\mu_2(\lambda_{\rm S})$.}
\centering
\label{figure2}
\end{figure}

\begin{figure*}[t]
\centerline{\includegraphics[width=\linewidth]{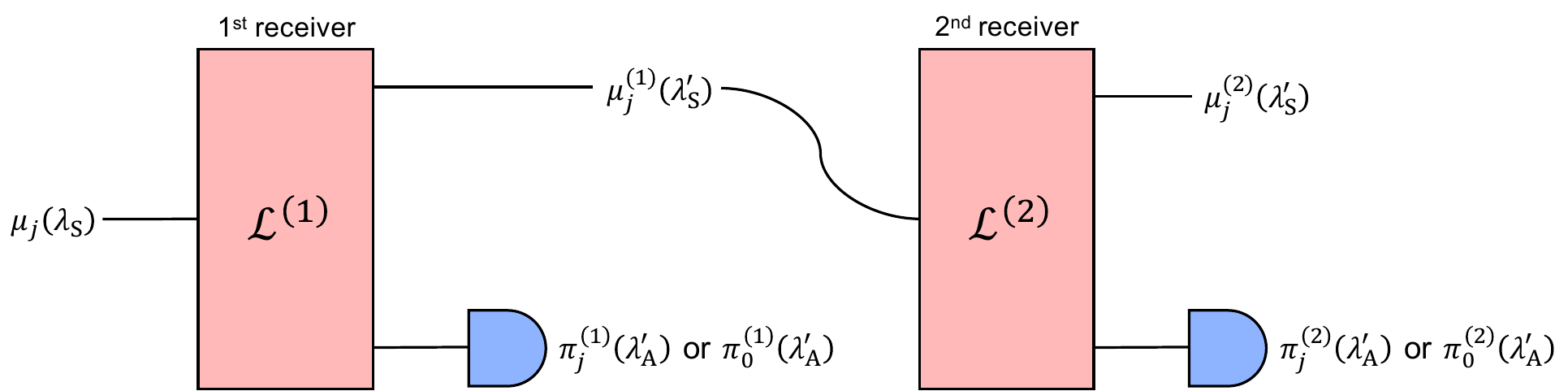}}
\caption{Non-demolition measurement reproduced within the noncontextual model for sequential unambiguous discrimination between two epistemic states $\mu_1(\lambda_{\rm S})$ and $\mu_2(\lambda_{\rm S})$, performed by two consecutive receivers.}
\centering
\label{figure3}
\end{figure*}

\subsection{Optimal unambiguous state discrimination}
We quantitatively analyze the capability of non-demolition measurements in several discrimination tasks. We first consider unambiguous state discrimination, in which two epistemic states $\mu_1(\lambda_{\rm S})$ and $\mu_2(\lambda_{\rm S})$ are discriminated without error. To this end, a non-demolition measurement can be constructed using the following transformation, as illustrated in Fig.~\ref{figure2}:
\begin{align}\label{ud}
    &\int_{\Omega_{\rm S}} d\lambda_{\rm S} \, \mathcal{L}(\lambda_{\rm S}', \lambda_{\rm A}', \lambda_{\rm S}) \mu_j(\lambda_{\rm S}) \nonumber\\
    &= \widetilde{\mu}_j(\lambda_{\rm S}') \left\{ \alpha_j {\sigma}_j(\lambda_{\rm A}') + (1 - \alpha_j){\sigma}_0(\lambda_{\rm A}') \right\}.
\end{align}
Here, $\alpha_j \in [0,1]$ denotes the probability of obtaining outcome $j$. The state $\widetilde{\mu}_j(\lambda_{\rm S}')$ represents the post-measurement epistemic state, and ${\sigma}_k(\lambda_{\rm A}')$ with $k \in \{j,0\}$ are {epistemic states that are orthogonal to each other, perfectly distinguished by the direct measurement} as defined in Eq.~(\ref{dir}). When the outcome is $j \neq 0$, the prepared state $\mu_j(\lambda_{\rm S})$ can be identified without error; otherwise, the outcome corresponds to failure.

Given prior probabilities $q_j$ for $\mu_j(\lambda_{\rm S})$, the average success probability is
\begin{align}\label{p_su}
    P_{\rm succ}^{\rm (NC)}
    &= q_1 p(1|\mathcal{M}_{\rm nd}, \mu_1) + q_2 p(2|\mathcal{M}_{\rm nd}, \mu_2) \nonumber\\
    &= q_1 \alpha_1 + q_2 \alpha_2,
\end{align}
which coincides with the corresponding expression in the quantum model.

The conditions on $\alpha_1$ and $\alpha_2$ under which $\mathcal{L}$ in Eq.~(\ref{ud}) is physically valid follow from the confusability-preserving condition in Eq.~(\ref{con_p}) (see Appendix~B.1 for details):
\begin{align}\label{rel_ud}
    \underbrace{\Big(\mu_1(\lambda_{\rm S}),\mu_2(\lambda_{\rm S})\Big)_{\Omega_{\rm S}}}_{=c}=\underbrace{\Big(\widetilde{\mu}_1(\lambda_{\rm S}'),\widetilde{\mu}_2(\lambda_{\rm S}')\Big)_{\Omega_{\rm S}'}}_{=c'}(1-\alpha_k),
\end{align}
which can be rewritten as
\begin{align}
    \alpha_k = 1 - \frac{c}{c'}.
\end{align}

Since $\alpha_k \in [0,1]$, it follows that $c' \ge c$. This suggests that it is physically valid to reduce the region of $\alpha_k$ to $[0,1-c]$, making a right rectangular region $[0,1-c]\times[0,1-c]$. Furthermore, all response functions derived from Eq.~(\ref{ud}) correspond to points $(\alpha_1, \alpha_2)$ lying below a straight line with intercepts in $[0,1-c]$, as shown in Appendix~B.2. Consequently, the physically allowed region is given by
\begin{align}\label{cond_nc}
    &\alpha_k \ge 0, \ \forall k \in \{1,2\}, \nonumber\\
    &\alpha_1 + \alpha_2 \le 1 - c.
\end{align}

Maximizing Eq.~(\ref{p_su}) over all $(\alpha_1, \alpha_2)$ satisfying Eq.~(\ref{cond_nc}) yields
\begin{eqnarray}
    \max_{\mathcal{M}_{\rm nd}} P_{\rm succ}^{\rm (NC)}=\max_{(\alpha_1,\alpha_2)} P_{\rm succ}^{\rm (NC)}=\max\{q_1,q_2\}(1-c),
\end{eqnarray}
recovering the result of previous work~\cite{j.shin2}.

By contrast, when $(\alpha_1, \alpha_2)$ correspond to a quantum non-demolition measurement, they must satisfy~\cite{m.namkung_anal}
\begin{align}\label{cond_q}
    &\alpha_k \ge 0, \ \forall k \in \{1,2\}, \nonumber\\
    &(1 - \alpha_1)(1 - \alpha_2) \ge c.
\end{align}
This leads to a maximum success probability that exceeds $\max_{(\alpha_1,\alpha_2)} P_{\rm succ}^{\rm (NC)}$ when the prior probabilities are sufficiently close ($|q_1 - q_2|$ small), and does not otherwise~\cite{g.jaeger}.

Both Eqs.~(\ref{cond_nc}) and (\ref{cond_q}) indicate that the noncontextual framework imposes stronger constraints on the structure of non-demolition measurements than the quantum model. Consequently, although the noncontextual model can reproduce the structure of non-demolition measurements, as in Eq.~(\ref{ud}), it fails to fully replicate discrimination tasks in which states are identified without error.

\subsection{Sequential unambiguous state discrimination}

We remark that non-demolition measurements do not completely destroy the information contained in the initial state. This implies that the remaining information can be accessed by a subsequent receiver who also performs unambiguous state discrimination, a scenario referred to as sequential unambiguous state discrimination~\cite{j.a.bergou_seq}. For simplicity, we consider two consecutive receivers discriminating between two epistemic states $\mu_1(\lambda_{\rm S})$ and $\mu_2(\lambda_{\rm S})$, as shown in Fig.~\ref{figure3}. This scenario can be generalized to $N$ receivers.

The non-demolition measurement performed by the $r$-th receiver is described by
\begin{align}
    &\int_{\Omega_{\rm S}} d\lambda_{\rm S} \, \mathcal{L}^{(r)}(\lambda_{\rm S}', \lambda_{\rm A}', \lambda_{\rm S}) \mu_j(\lambda_{\rm S}) \nonumber\\
    &= \mu_j^{(r)}(\lambda_{\rm S}') \left\{ \alpha_j^{(r)} {\sigma}_j^{(r)}(\lambda_{\rm A}') + (1 - \alpha_j^{(r)}) {\sigma}_0^{(r)}(\lambda_{\rm A}') \right\}.
\end{align}
Here, $\alpha_j^{(r)}$ denotes the probability of obtaining outcome $j$, and $\{{\sigma}_k^{(r)}(\lambda_{\rm A}')\}_{k=0}^{2}$ describes the {orthogonal epistemic states}.

The probability that both receivers successfully identify the state is
\begin{align}
    P_{\rm succ}^{\rm (NC)}
    &= \sum_{k\in\{1,2\}} q_k \, p(k|\mathcal{M}_{\rm nd}^{(1)}, \mu_k) \, p(k|\mathcal{M}_{\rm nd}^{(2)}, \mu_k^{(1)}) \nonumber\\
    &= q_1 \alpha_1^{(1)} \alpha_1^{(2)} + q_2 \alpha_2^{(1)} \alpha_2^{(2)}.
\end{align}

This quantity is maximized over $\alpha_j^{(r)} \ge 0$ subject to
\begin{align}
    \alpha_1^{(1)} + \alpha_2^{(1)} &\le 1 - c, \nonumber\\
    \alpha_1^{(2)} + \alpha_2^{(2)} &\le 1 - c^{(1)},
\end{align}
where $c^{(1)}$ denotes the confusability between $\mu_1^{(1)}(\lambda_{\rm S}')$ and $\mu_2^{(1)}(\lambda_{\rm S}')$.

The maximum success probability is obtained as
\begin{eqnarray}\label{succ_seq}
    \max_{\mathcal{M}_{\rm nd}^{(1)},\mathcal{M}_{\rm nd}^{(2)}}P_{\rm succ}^{\rm (NC)}=\max\{q_1,q_2\}(1-\sqrt{c})^{2},
\end{eqnarray}
as derived in Appendix~C. This coincides with the case in which only a single quantum state is discriminated using a quantum non-demolition measurement.

For comparison, the maximum success probability in the quantum model under equal priors is~\cite{j.a.bergou_seq,c.-q.pang}
\begin{align}
    \max P_{\rm succ}^{\rm (Q)} =
    \begin{cases}
        (1 - c^{1/4})^2, & c < \sqrt{3 - 2\sqrt{2}},\\
        \frac{1}{2}(1 - \sqrt{c})^2, & c \ge \sqrt{3 - 2\sqrt{2}}.
    \end{cases}
\end{align}

This shows that as $c$ increases, the contextual advantage in sequential unambiguous state discrimination diminishes. In particular, the noncontextual model fails to reproduce the quantum result only in the regime of small confusability.

When the number of receivers is generalized to $N$, the maximum success probability becomes
\begin{align}
    \max_{\mathcal{M}_{\rm nd}^{(1)}, \ldots, \mathcal{M}_{\rm nd}^{(N)}} P_{\rm succ}^{\rm (NC)}
    = \max\{q_1, q_2\} (1 - c^{1/N})^N,
\end{align}
which again coincides with the case of discriminating a single quantum state in the quantum model.

We compare with the quantum model providing the maximum success probability~\cite{m.namkung_gen}
\begin{align}
    \max P_{\rm succ}^{\rm (Q)} =
    \begin{cases}
        (1 - c^{1/2N})^N, & c < (2^{1/N} - 1)^{N/2},\\
        \frac{1}{2}(1 - c^{1/N})^N, & c \ge (2^{1/N} - 1)^{N/2},
    \end{cases}
\end{align}
under equal priors. It can be verified that the region in which $\frac{1}{2}(1-{c}^{1/N})^N$ becomes the maximum increases with $N$. This suggests that, when the number of the receivers becomes larger, the maximum success probability tends to be reproduced by the noncontextual model.

\begin{figure*}[t]
\centerline{\includegraphics[width=\linewidth]{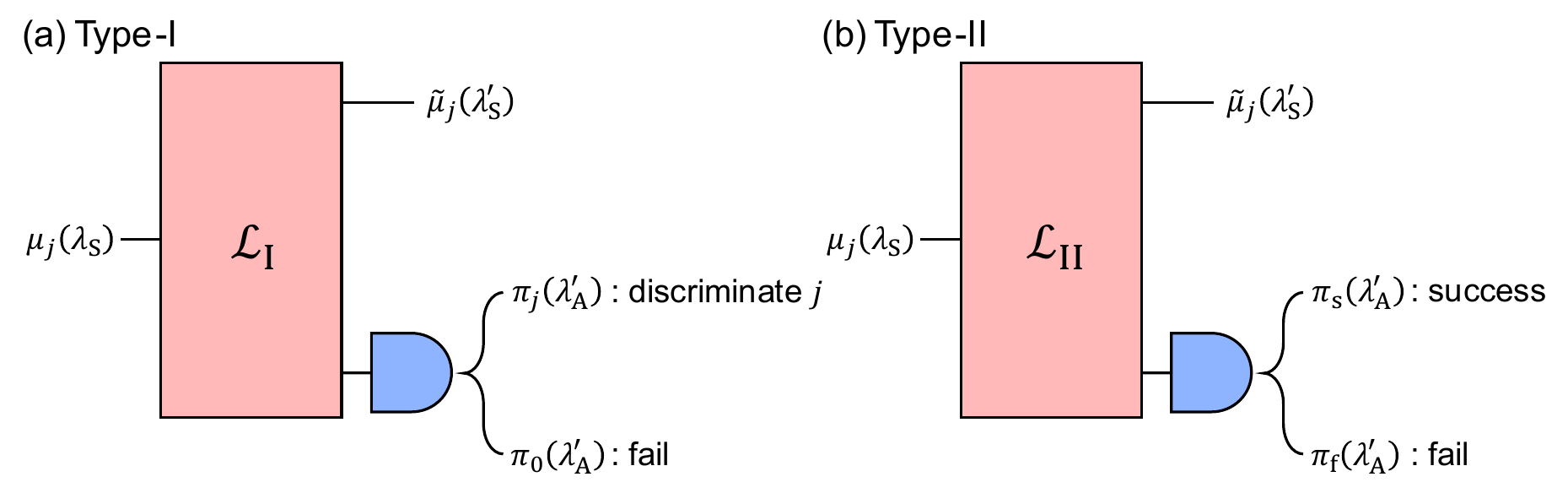}}
\caption{Non-demolition measurement reproduced within the noncontextual model for probabilistic quantum cloning. (a) Type-I probabilistic quantum cloning, in which two epistemic states are also unambiguously discriminated. (b) Type-II probabilistic quantum cloning, in which only quantum cloning is performed with a certain probability.}
\centering
\label{figure4}
\end{figure*}

\subsection{Probabilistic quantum cloning}

We further analyze the contextual advantage of probabilistic quantum cloning, which produces perfect clones with a certain probability. We first consider type-I probabilistic cloning, in which the information about the prepared state is also revealed to the receiver, followed by type-II probabilistic cloning~\cite{l.m.duan_cl}, where only the success of cloning is of interest.

\subsubsection{Type-I probabilistic cloning}
Type-I probabilistic cloning, illustrated in Fig.~\ref{figure4}(a), is implemented via a transformation $\mathcal{L}_{\rm I}$ such that
\begin{align}\label{t1c}
    &\int_{\Omega_{\rm S}} d\lambda_{\rm S} \, \mathcal{L}_{\rm I}(\lambda_{\rm S1}', \lambda_{\rm S2}', \lambda_{\rm A}', \lambda_{\rm S}) \mu_j(\lambda_{\rm S}) \nonumber\\
    &= \alpha_j \, \mu_j(\lambda_{\rm S1}') \mu_j(\lambda_{\rm S2}') {\sigma}_j(\lambda_{\rm A}') \nonumber\\
    &\quad + (1 - \alpha_j) \, \chi_j(\lambda_{\rm S1}', \lambda_{\rm S2}') {\sigma}_0(\lambda_{\rm A}'),
\end{align}
followed by a direct measurement on the auxiliary system, described by {orthogonal epistemic states} ${\sigma}_0(\lambda_{\rm A}')$, ${\sigma}_1(\lambda_{\rm A}')$, and ${\sigma}_2(\lambda_{\rm A}')$.

Here, $\chi_j(\lambda_{\rm S1}', \lambda_{\rm S2}')$ denotes the post-measurement epistemic state on $\Omega_{\rm S1}' \times \Omega_{\rm S2}'$ corresponding to the failure outcome $j=0$. When the measurement outcome is nonzero, the epistemic state $\mu_j(\lambda_{\rm S})$ is perfectly cloned into $\mu_j(\lambda_{\rm S1}')$ and $\mu_j(\lambda_{\rm S2}')$, while simultaneously being unambiguously discriminated.

Applying the confusability-preserving condition of Eq.~(\ref{con_p}) to Eq.~(\ref{t1c}), we obtain the constraint
\begin{align}
    c=(1-\alpha_k)\Big(\chi_1(\lambda_{\rm S1}',\lambda_{S2}'),\chi_2(\lambda_{\rm S1}',\lambda_{S2}')\Big)_{\Omega_{\rm S1}\times\Omega_{\rm S2}},
\end{align}
where $c$ denotes the confusability between $\mu_1(\lambda_{\rm S})$ and $\mu_2(\lambda_{\rm S})$. This relation can be rewritten as
\begin{eqnarray}
    \alpha_k&=&1-\frac{c}{\Big(\chi_1(\lambda_{\rm S1}',\lambda_{S2}'),\chi_2(\lambda_{\rm S1}',\lambda_{S2}')\Big)_{\Omega_{\rm S1}\times\Omega_{\rm S2}}}\nonumber\\
    &\le&1-c.
\end{eqnarray}

Since the transformation $\mathcal{L}_{\rm I}$ in Eq.~(\ref{t1c}) has the same structure as $\mathcal{L}$ in Eq.~(\ref{ud}), the maximum success probability of type-I probabilistic cloning is readily obtained as
\begin{align}
    \max_{\mathrm{type}\text{-}\mathrm{I}} P_{\rm succ}^{\rm (NC)} = \max\{q_1, q_2\} (1 - c),
\end{align}
which is generally lower than the corresponding maximum success probability in the quantum model, given by the IDP limit~\cite{i.d.ivanovic,d.dieks,a.peres,g.jaeger}.

\subsubsection{Type-II probabilistic cloning}

Type-II probabilistic cloning, illustrated in Fig.~\ref{figure4}(b) and originally proposed by Duan and Guo~\cite{l.m.duan_cl}, is described by a transformation $\mathcal{L}_{\rm II}$ such that
\begin{align}\label{t2c}
    &\int_{\Omega_{\rm S}} d\lambda_{\rm S} \, \mathcal{L}_{\rm II}(\lambda_{\rm S1}', \lambda_{\rm S2}', \lambda_{\rm A}', \lambda_{\rm S}) \mu_j(\lambda_{\rm S}) \nonumber\\
    &= \alpha_j \, \mu_j(\lambda_{\rm S1}') \mu_j(\lambda_{\rm S2}') {\sigma}_{\rm s}(\lambda_{\rm A}') \nonumber\\
    &\quad + (1 - \alpha_j) \, \chi_j(\lambda_{\rm S1}', \lambda_{\rm S2}') {\sigma}_{\rm f}(\lambda_{\rm A}'),
\end{align}
followed by {orthogonal epistemic states} ${\sigma}_{\rm s}(\lambda_{\rm A}')$ and ${\sigma}_{\rm f}(\lambda_{\rm A}')$, corresponding to success and failure outcomes, respectively.

The transformation $\mathcal{L}_{\rm II}$ satisfies the confusability-preserving condition if the following relation holds (see Appendix~D.1 for details):
\begin{align}\label{pc_cond}
    c&=\alpha_kc^2+(1-\alpha_k)\Big(\chi_1(\lambda_{\rm S1}',\lambda_{S2}'),\chi_2(\lambda_{\rm S1}',\lambda_{S2}')\Big)\nonumber\\
    &\le\alpha_kc^2+1-\alpha_k,
\end{align}
which can be rewritten as
\begin{align}
    \alpha_k \le \frac{1}{1 + c}.
\end{align}

Since $\alpha_k$ denotes the success probability of cloning for a given input $\mu_k(\lambda_{\rm S})$, this yields the maximum success probability
\begin{align}
    \max_{\mathrm{type}\text{-}\mathrm{II}} P_{\rm succ}^{\rm (NC)} = \max\{q_1, q_2\} \frac{1}{1 + c},
\end{align}
for given prior probabilities $q_1$ and $q_2$.

This analysis can be extended to the case of generating $m > n$ copies from $n$ initial copies, leading to
\begin{align}\label{gen_pr_cl}
    \max_{\mathrm{type}\text{-}\mathrm{II}} P_{\rm succ}^{\rm (NC)} = \max\{q_1, q_2\} \frac{1 - c^n}{1 - c^m},
\end{align}
as shown in Appendix~D.2.

For comparison, the corresponding quantum bound is given by~\cite{w.-h.zhang}
\begin{align}
    \max_{\mathrm{type}\text{-}\mathrm{II}} P_{\rm succ}^{\rm (Q)} = \frac{1 - (\sqrt{c})^n}{1 - (\sqrt{c})^m}.
\end{align}
This indicates that, for equal prior probabilities, the success probability achievable within the noncontextual model exceeds that of the quantum model for all $m > n$. This gap implies contextuality in terms of cloning success probability.

On the other hand, when the prior probabilities become highly imbalanced, the quantum model approaches a noncontextual description. For example, when $q_1 \gg q_2$ (or vice versa), the maximum success probability in the quantum model becomes
\begin{align}\label{cl_qm}
    \max_{\mathrm{type}\text{-}\mathrm{II}} P_{\rm succ}^{\rm (Q)} = \max\{q_1, q_2\}(1 - c^n),
\end{align}
as derived in Appendix~D.3, which remains strictly smaller than Eq.~(\ref{gen_pr_cl}) for $m > n$.

\subsection{Extension to noisy state discrimination}

So far, we have theoretically demonstrated contextual features of quantum non-demolition measurements in discrimination tasks, primarily in the noiseless case. We now extend our analysis to noisy quantum states, leading to the scenario of maximal-confidence discrimination.

In the presence of noise, the maximum confidence for outcome $k$, defined as~\cite{s.croke,k.flatt}
\begin{equation}\label{cnc}
    C^{\rm (NC)}(k)=\max_{\xi_k(\lambda)}\frac{\int_{\Omega_{\rm S}} d\lambda_{\rm S}\, q_k \mu_k(\lambda_{\rm S}) \xi_k(\lambda_{\rm S})}{\int_{\Omega_{\rm S}} d\lambda_{\rm S}\, \mu(\lambda_{\rm S}) \xi_k(\lambda_{\rm S})},
\end{equation}
does not generally attain unity, where $\mu(\lambda_{\rm S}) = q_1 \mu_1(\lambda_{\rm S}) + q_2 \mu_2(\lambda_{\rm S})$.

Defining two functions
\begin{align}
    \zeta_k(\lambda_{\rm S}) &= q_k \mu_k(\lambda_{\rm S}) \mu^{-1}(\lambda_{\rm S}), \nonumber\\
    Q_k(\lambda_{\rm S}) &= \frac{\mu(\lambda_{\rm S}) \xi_k(\lambda_{\rm S})}{\int_{\Omega_{\rm S}} d\lambda_{\rm S}\, \mu(\lambda_{\rm S}) \xi_k(\lambda_{\rm S})},
\end{align}
the maximal confidence in Eq.~(\ref{cnc}) can be rewritten as
\begin{align}\label{max_conf}
    C^{\rm (NC)}(k) = \max_{\substack{Q_k(\lambda_{\rm S}) \ge 0 \\ \int_{\Omega_{\rm S}} d\lambda_{\rm S}\, Q_k(\lambda_{\rm S}) = 1}} \int_{\Omega_{\rm S}} d\lambda_{\rm S}\, \zeta_k(\lambda_{\rm S}) Q_k(\lambda_{\rm S}).
\end{align}

Since evaluating the maximal confidence reduces to a convex optimization problem (see Appendix~E.1)~\cite{s.boyd}, it can equivalently be obtained by solving the dual problem:
\begin{align}
    \text{minimize} \quad & l_k, \nonumber\\
    \text{subject to} \quad & l_k \mu(\lambda_{\rm S}) - q_k \mu_k(\lambda_{\rm S}) \ge 0, \ \forall \lambda_{\rm S} \in \Omega_{\rm S}.
\end{align}

For the optimal value of the dual problem to coincide with Eq.~(\ref{max_conf}), the response functions $\xi_k(\lambda_{\rm S})$ must satisfy the complementary slackness condition. This implies that the corresponding measurement unambiguously discriminates complementary states $\sigma_k(\lambda_{\rm S})$ associated with $\mu_k(\lambda_{\rm S})$~\cite{j.bae,h.lee_mc}.

This observation provides a physical interpretation of contextuality-enhanced maximal-confidence discrimination: the enhancement arises because a quantum non-demolition measurement that unambiguously discriminates complementary states cannot, in general, be reproduced within a noncontextual model. 

This insight is particularly relevant for experimental implementations, where maximal-confidence discrimination can be realized via linear optical setups designed to unambiguously discriminate complementary states (see Appendix~E.2 for details).

\begin{table}[t]
    \centering
    \begin{tabular}{c|c|c|c}
    \hline
      Task & Priors & Contextual regime & Noncontextual regime \\
      \hline \hline 
       USD~\cite{j.shin2} & Equal  & Always & -- \\
       USD~\cite{j.shin2} & Unequal  & Small $\Delta q$ and $c$ & Large $\Delta q$ and $c$ \\
       \hline
       SUSD & Equal & Small $N$ and $c$ & Large $N$ and $c$ \\
       \hline
       PQC-I & Equal & Always & -- \\
       PQC-I & Unequal & Small $c$ & Large $c$ \\
       \hline
       PQC-II & Equal & Always & -- \\
       PQC-II & Unequal & Small $\Delta q$ & Large $\Delta q$ \\
       \hline
    \end{tabular}
    \caption{Summary of contextuality enhancement in unambiguous state discrimination (USD), sequential unambiguous state discrimination (SUSD), and type-I and type-II probabilistic quantum cloning (PQC-I and PQC-II, respectively). Here, $\Delta q = |q_1 - q_2|$, while $c$ and $N$ denote the confusability and the number of receivers, respectively. The USD results are included for comparison with previous work~\cite{j.shin2}, demonstrating that our framework recovers known results.}
    \label{table1}
\end{table}

\section{Conclusion}

We rigorously verified the contextual aspects of quantum non-demolition measurements used in various discrimination tasks. We first formulated a framework for non-demolition measurements within a noncontextual model, in terms of state preparation, transformation, and measurement, as established in previous works~\cite{r.w.spekkens,d.schmid}. Based on this framework, all discrimination tasks considered in this work are \textit{analytically} characterized, as summarized in Table~\ref{table1}. 

For unambiguous state discrimination, contextuality enhancement is always (conditionally) observed under equal (unequal) prior probabilities, consistent with the predictions of our methodology. In sequential unambiguous state discrimination, contextuality tends to decrease as the number of receivers and the degree of confusability increase, even when the two states are equiprobable. We further classify probabilistic quantum cloning into two types.  In type-I probabilistic cloning, contextuality exhibits the same behavior as in unambiguous state discrimination. Type-II cloning is also enhanced by contextuality under equal priors, while tending to admit a noncontextual description under unequal priors.

We believe that our methodology provides a unified approach to identifying quantum resources relevant for advancing various quantum technologies, as well as for deepening the understanding of the foundations of quantum measurements. Quantum non-demolition measurements are widely used in quantum communication tasks~\cite{r.han}, where security against eavesdroppers is ensured~\cite{m.namkung_qkd}. In particular, probabilistic quantum cloning can be generalized to probabilistic quantum broadcasting~\cite{l.li}, which distributes copied information to multiple parties via entanglement. Moreover, since quantum non-demolition measurements have been experimentally implemented in various setups~\cite{p.grangier,t.nakajima}, our method may pave the way for observing nonclassical features in experimental realizations.

\section*{Acknowledgements}
This work was partly supported by a National Research Foundation of Korea (NRF) grant funded by the Korea government (MSIT) (RS-2022-NR068817, RS-2023-NR119925, RS-2024-00336079), an Institute for Information \& Communications Technology Planning \& Evaluation (IITP) grant funded by the Korea government (MSIT) (RS-2025-02292999), and the KIST research program (26E0001, 26E0011).

\section*{Data availability}
The data are available from the authors upon reasonable request.

\begin{widetext}

\section*{Appendix A. Establishing response functions from the transformation formalism}

We first recall the measurement probability in Eq.~(\ref{meas_prob}) in order to derive the response functions:
\begin{align}\label{p_append}
    p(k|\mathcal{M}_{\rm nd}, \mu)
    &= \int_{\Omega_{\rm S}' \times \Omega_{\rm A}' \times \Omega_{\rm S}} d\lambda_{\rm S}' d\lambda_{\rm A}' d\lambda_{\rm S} \, \pi_k(\lambda_{\rm A}') \mathcal{L}(\lambda_{\rm S}', \lambda_{\rm A}', \lambda_{\rm S}) \mu(\lambda_{\rm S}) \nonumber\\
    &= \int_{\Omega_{\rm S}' \times \Omega_{\rm S}} d\lambda_{\rm S}' d\lambda_{\rm S} 
    \underbrace{\int_{\Omega_{\rm A}'} d\lambda_{\rm A}' \, \pi_k(\lambda_{\rm A}') \mathcal{L}(\lambda_{\rm S}', \lambda_{\rm A}', \lambda_{\rm S})}_{=\mathcal{L}_k(\lambda_{\rm S}', \lambda_{\rm S})}
    \mu(\lambda_{\rm S}) = \int_{\Omega_{\rm S}} d\lambda_{\rm S} \left\{ \int_{\Omega_{\rm S}'} d\lambda_{\rm S}' \, \mathcal{L}_k(\lambda_{\rm S}', \lambda_{\rm S}) \right\} \mu(\lambda_{\rm S}).
\end{align}

We define the functions $\xi_k(\lambda_{\rm S})$ as
\begin{align}
    \xi_k(\lambda_{\rm S}) = \int_{\Omega_{\rm S}'} d\lambda_{\rm S}' \, \mathcal{L}_k(\lambda_{\rm S}', \lambda_{\rm S}).
\end{align}
These functions are non-negative since $\mathcal{L}_k(\lambda_{\rm S}', \lambda_{\rm S}) \ge 0$ for all $(\lambda_{\rm S}', \lambda_{\rm S}) \in \Omega_{\rm S}' \times \Omega_{\rm S}$.

Moreover, summing $\xi_k(\lambda_{\rm S})$ over all $k$ yields
\begin{align}
    \sum_k \xi_k(\lambda_{\rm S})
    &= \sum_k \int_{\Omega_{\rm S}'} d\lambda_{\rm S}' \, \mathcal{L}_k(\lambda_{\rm S}', \lambda_{\rm S})= \int_{\Omega_{\rm S}' \times \Omega_{\rm A}'} d\lambda_{\rm S}' d\lambda_{\rm A}' 
    \left\{ \sum_k \pi_k(\lambda_{\rm A}') \right\}
    \mathcal{L}(\lambda_{\rm S}', \lambda_{\rm A}', \lambda_{\rm S}) \nonumber\\
    &= \int_{\Omega_{\rm S}' \times \Omega_{\rm A}'} d\lambda_{\rm S}' d\lambda_{\rm A}' 
    \mathcal{L}(\lambda_{\rm S}', \lambda_{\rm A}', \lambda_{\rm S})= 1,
\end{align}
where the third equality follows from the normalization condition of the response functions $\pi_k(\lambda_{\rm A}')$, and the final equality follows from Eq.~(\ref{LL}).

    \section*{Appendix B: non-demolition measurement for unambiguous state discrimination}

    \subsection*{B.1. Verification of Eqs.~(\ref{p_su}) and (\ref{rel_ud})}
    Here we analytically verify Eqs.~(\ref{p_su}) and (\ref{rel_ud}). First, the measurement probability $p(j|\mathcal{M}_{\rm nd},\mu_j)$ is evaluated as
    \begin{align}
        p(j|\mathcal{M}_{\rm nd},\mu_j)&=\int_{\Omega_{\rm S}'\times\Omega_{\rm A}'}d\lambda_{\rm S}'d\lambda_{\rm A}'\pi_j(\lambda_{\rm A}')\int_{\Omega_{\rm S}}d\lambda_{\rm S}\mathcal{L}(\lambda_{\rm S}',\lambda_{\rm A}',\lambda_{\rm S})\mu_j(\lambda_{\rm S})\\
        &=\int_{\Omega_{\rm S}'\times\Omega_{\rm A}'}d\lambda_{\rm S}'d\lambda_{\rm A}'\pi_j(\lambda_{\rm A}')\widetilde{\mu}_j(\lambda_{\rm S}')\left\{\alpha_j{\sigma}_j(\lambda_{\rm A}')+(1-\alpha_j){\sigma}_0(\lambda_{\rm A}')\right\}\nonumber\\
        &=\alpha_j\int_{\Omega_{\rm S}'}d\lambda_{\rm S}'\widetilde{\mu}_j(\lambda_{\rm S}')\int_{\Omega_{\rm A}'}d\lambda_{A}'\pi_j(\lambda_{\rm A}'){\sigma_j(\lambda_{\rm A}')}=\alpha_j,\nonumber
    \end{align}
    leading to the success probability of Eq.~(\ref{p_su}). 

To evaluate the confusability-preserving condition of Eq.~(\ref{rel_ud}), we remind that the support space of $\int_{\Omega_{\rm S}}d\lambda_{\rm S}\mathcal{L}(\lambda_{\rm S}',\lambda_{\rm A}',\lambda_{\rm S})\mu_j(\lambda_{\rm S})$ is
    \begin{align}
        \mathrm{supp}\left[\int_{\Omega_{\rm S}}d\lambda_{\rm S}\mathcal{L}(\lambda_{\rm S}',\lambda_{\rm A}',\lambda_{\rm S})\mu_j(\lambda_{\rm S})\right]&=\mathrm{supp}\left[\widetilde{\mu}_j(\lambda_{\rm S}')\left\{\alpha_j{\sigma}_j(\lambda_{\rm A}')+(1-\alpha_j){\sigma}_0(\lambda_{\rm A}')\right\}\right]\nonumber\\
        &=\mathrm{supp}\left[\widetilde{\mu}_j(\lambda_{\rm S}')\right]\times\mathrm{supp}\left[\alpha_j{\sigma}_j(\lambda_{\rm A}')+(1-\alpha_j){\sigma}_0(\lambda_{\rm A}')\right]\nonumber\\
        &=\mathrm{supp}\left[\widetilde{\mu}_j(\lambda_{\rm S}')\right]\times\left\{\mathrm{supp}[{\sigma}_j(\lambda_{\rm A}')]\cup\mathrm{supp}[{\sigma}_0(\lambda_{\rm A}')]\right\},
    \end{align}
    where ``$\times$'' denotes Cartesian product. Therefore, $\Big(\mu_1(\lambda_{\rm S}),\mu_2(\lambda_{\rm S})\Big)_{\Omega_{\rm S}}$ in Eq.~(\ref{rel_ud}) is evaluated as
    \begin{align}
        \Big(\mu_1(\lambda_{\rm S}),\mu_2(\lambda_{\rm S})\Big)_{\Omega_{\rm S}}&=\Big(\int_{\Omega_{\rm S}}d\lambda_{\rm S}\mathcal{L}(\lambda_{\rm S}',\lambda_{\rm A}',\lambda_{\rm S})\mu_1(\lambda_{\rm S}),\int_{\Omega_{\rm S}}d\lambda_{\rm S}\mathcal{L}(\lambda_{\rm S}',\lambda_{\rm A}',\lambda_{\rm S})\mu_2(\lambda_{\rm S})\Big)_{\Omega_{\rm S}'\times\Omega_{\rm A}'}\nonumber\\
        &=\int_{\mathrm{supp}\left[\int_{\Omega_{\rm S}}d\lambda_{\rm S}\mathcal{L}(\lambda_{\rm S}',\lambda_{\rm A}',\lambda_{\rm S})\mu_1(\lambda_{\rm S})\right]}d\lambda_{\rm S}'d\lambda_{\rm A}'\left\{\int_{\Omega_{\rm S}}d\lambda_{\rm S}\mathcal{L}(\lambda_{\rm S}',\lambda_{\rm A}',\lambda_{\rm S})\mu_2(\lambda_{\rm S})\right\}\nonumber\\
        &=\int_{\mathrm{supp}\left[\widetilde{\mu}_1(\lambda_{\rm S}')\right]\times\left\{\mathrm{supp}[{\sigma}_1(\lambda_{\rm A}')]\cup\mathrm{supp}[{\sigma}_0(\lambda_{\rm A}')]\right\}}d\lambda_{\rm S}'d\lambda_{\rm A}'\widetilde{\mu}_2(\lambda_{\rm S}')\left\{\alpha_2{\sigma}_2(\lambda_{\rm A}')+(1-\alpha_2){\sigma}_0(\lambda_{\rm A}')\right\}\nonumber\\
        &=\int_{\mathrm{supp[\widetilde{\mu}_1(\lambda_{\rm S}')]}}d\lambda_{\rm S}'\widetilde{\mu}_2(\lambda_{\rm S}')\int_{\mathrm{supp}[{\sigma}_1(\lambda_{\rm A}')]\cup\mathrm{supp}[{\sigma}_0(\lambda_{\rm A}')]}d\lambda_{\rm A}'\left\{\alpha_2{\sigma}_2(\lambda_{\rm A}')+(1-\alpha_2){\sigma}_0(\lambda_{\rm A}')\right\}\nonumber\\
        &=\Big(\mu_1(\lambda_{\rm S}'),\mu_2(\lambda_{\rm S}')\Big)_{\Omega_{\rm S}'}(1-\alpha_2).
    \end{align}
    Likewise, we can obtain the relation $\Big(\mu_1(\lambda_{\rm S}),\mu_2(\lambda_{\rm S})\Big)_{\Omega_{\rm S}}=\Big(\mu_1(\lambda_{\rm S}'),\mu_2(\lambda_{\rm S}')\Big)_{\Omega_{\rm S}'}(1-\alpha_1)$, thereby finishing the verification.

\subsection*{B.2. Derivation of the constraint in Eq.~(\ref{cond_nc})}

We begin by expressing the transformation as
\begin{align}
    \mathcal{L}_k(\lambda_{\rm S}', \lambda_{\rm A}', \lambda_{\rm S})
    ={\alpha_k} \widetilde{\mu}_k(\lambda_{\rm S}')
     {\sigma}_k(\lambda_{\rm A}') 
    \bar{\mu}_k(\lambda_{\rm S}),
    \quad k \neq 0,
\end{align}
such that $\sum_k \mathcal{L}_k{(\lambda_{\rm S}', \lambda_{\rm A}', \lambda_{\rm S})} = \mathcal{L}(\lambda_{\rm S}', \lambda_{\rm A}', \lambda_{\rm S})$.

Here, $\bar{\mu}_k(\lambda_{\rm S})$ satisfies the orthogonality condition
\begin{align}
    \int_{\Omega_{\rm S}} d\lambda_{\rm S} \, \bar{\mu}_k(\lambda_{\rm S}) \mu_l(\lambda_{\rm S}) = \delta_{kl},
\end{align}
with $\delta_{kl}$ denoting the Kronecker delta.

Following the construction in Appendix~A, the response functions $\xi_k(\lambda_{\rm S})$ for $k=1,2$ are given by
\begin{align}
    \xi_k(\lambda_{\rm S}) = \alpha_k \bar{\mu}_k(\lambda_{\rm S}).
\end{align}

The completeness condition of the response functions then implies
\begin{align}
    \xi_0(\lambda_{\rm S})
    &= 1 - \xi_1(\lambda_{\rm S}) - \xi_2(\lambda_{\rm S}) \nonumber\\
    &= 1 - \alpha_1 \bar{\mu}_1(\lambda_{\rm S}) - \alpha_2 \bar{\mu}_2(\lambda_{\rm S}) \ge 0,
\end{align}
which is equivalent to
\begin{align}
    \alpha_1 \bar{\mu}_1(\lambda_{\rm S}) + \alpha_2 \bar{\mu}_2(\lambda_{\rm S}) \le 1.
\end{align}

This constraint defines a convex region in the $(\alpha_1, \alpha_2)$ plane. In particular, its extremal points are given by $(1 - c, 0)$ and $(0, 1 - c)$, leading to the condition in Eq.~(\ref{cond_nc}).

\section*{Appendix C: Derivation of the maximum success probability in Eq.~(\ref{succ_seq})}

We analytically derive the maximum success probability for sequential unambiguous state discrimination within the noncontextual model.

Using linear programming, the optimal pair $(\alpha_1^{(2)}, \alpha_2^{(2)})$ that maximizes the success probability is given by one of the extremal points $(1 - c', 0)$ or $(0, 1 - c')$. Therefore, the success probability satisfies
\begin{align}
    P_{\rm succ}^{\rm (NC)}
    \le \max \left\{
    q_1 \alpha_1 (1 - c'), \,
    q_2 \alpha_2 (1 - c')
    \right\}.
\end{align}

Using the relation $c' = \frac{c}{1 - \alpha_k}$, this can be rewritten as
\begin{align}
    P_{\rm succ}^{\rm (NC)}
    \le \max \left\{
    q_1 \alpha_1 \left(1 - \frac{c}{1 - \alpha_1}\right), \,
    q_2 \alpha_2 \left(1 - \frac{c}{1 - \alpha_2}\right)
    \right\}.
\end{align}

We now maximize the function
\begin{align}
    f(\alpha_k) = \alpha_k \left(1 - \frac{c}{1 - \alpha_k}\right),
\end{align}
over the interval $\alpha_k \in [0, 1 - c]$. A direct optimization yields
\begin{align}
    \max_{\alpha_k \in [0, 1 - c]} f(\alpha_k) = (1 - \sqrt{c})^2.
\end{align}

Substituting this result, we obtain
\begin{align}
    P_{\rm succ}^{\rm (NC)}
    \le \max\{q_1, q_2\} (1 - \sqrt{c})^2,
\end{align}
which establishes Eq.~(\ref{succ_seq}).

    \section*{Appendix D: non-demolition measurement for type-II probabilistic cloning}

    \subsection*{D.1. Verification of Eq.~(\ref{pc_cond})}
    We discuss the structure of the non-demolition measuremenet that performs the type-II probabilistic cloning. Note that the verification of the type-I cloning can also be performed in the same way. From the confusability-preserving condition, we obtain
    \begin{align}
        &\Big(\mu_1(\lambda_{\rm S}),\mu_2(\lambda_{\rm S})\Big)_{\Omega_{\rm S}}=\Big(\int_{\Omega_{\rm S}}d\lambda_{\rm S}\mathcal{L}^{(\rm II)}(\lambda_{\rm S1}',\lambda_{\rm S2}',\lambda_{\rm A}',\lambda_{\rm S})\mu_1(\lambda_{\rm S}),\int_{\Omega_{\rm S}}d\lambda_{\rm S}\mathcal{L}^{(\rm II)}(\lambda_{\rm S1}',\lambda_{\rm S2}',\lambda_{\rm A}',\lambda_{\rm S})\mu_2(\lambda_{\rm S})\Big)_{\Omega_{\rm S1}'\times\lambda_{\rm S2}'\times\Omega_{\rm A}'}\nonumber\\
        &=\int\limits_{\substack{\mathrm{supp}\big[\alpha_1\mu_1(\lambda_{\rm S1}')\mu_1(\lambda_{\rm S2}'){\sigma}_{\rm s}(\lambda_{\rm A}')\\+(1-\alpha_1)\chi_1(\lambda_{\rm S1}',\lambda_{\rm S2}'){\sigma}_{\rm f}(\lambda_{\rm A}')\big]}}d\lambda_{\rm S1}'d\lambda_{\rm S2}'d\lambda_{\rm A}'\left\{\alpha_2\mu_2(\lambda_{\rm S1}')\mu_2(\lambda_{\rm S2}'){\sigma}_{\rm s}(\lambda_{\rm A}')+(1-\alpha_2)\chi_2(\lambda_{\rm S1}',\lambda_{\rm S2}'){\sigma}_{\rm f}(\lambda_{\rm A}')\right\}\nonumber\\
        &=\int\limits_{\substack{\left\{\mathrm{supp}[\mu_1(\lambda_{\rm S1}')]\times\mathrm{supp}[\mu_1(\lambda_{\rm S2}')]\times\mathrm{supp}[{\sigma}_{\rm s}(\lambda_{\rm A}')]\right\}
        \\ \cup\left\{\mathrm{supp}[\chi_1(\lambda_{\rm S1}',\lambda_{\rm S2}')]\times\mathrm{supp}[{\sigma}_{\rm f}(\lambda_{\rm A}')]\right\}}}d\lambda_{\rm S1}'d\lambda_{\rm S2}'d\lambda_{\rm A}'\left\{\alpha_2\mu_2(\lambda_{\rm S1}')\mu_2(\lambda_{\rm S2}'){\sigma}_{\rm s}(\lambda_{\rm A}')+(1-\alpha_2)\chi_2(\lambda_{\rm S1}',\lambda_{\rm S2}'){\sigma}_{\rm f}(\lambda_{\rm A}')\right\}\nonumber\\
        &=\alpha_2\int_{\mathrm{supp}[\mu_1(\lambda_{\rm S1}')]}d\lambda_{\rm S1}'\mu_2(\lambda_{\rm S1}')\int_{\mathrm{supp}[\mu_1(\lambda_{\rm S2}')]}d\lambda_{\rm S2}'\mu_2(\lambda_{\rm S2}')+(1-\alpha_2)\int_{\mathrm{supp}[\chi_1(\lambda_{\rm S1}',\lambda_{\rm S2}')]}d\lambda_{\rm S1}'\lambda_{\rm S2}'\chi_2(\lambda_{\rm S1}',\lambda_{\rm S2}')\nonumber\\
        &\le\alpha_2\Big(\mu_1(\lambda_{\rm S}),\mu_2(\lambda_{\rm S})\Big)_{\Omega_{\rm S}}^2+1-\alpha_2.
    \end{align}
    In particular, we use that there is no intersection between $\mathrm{supp}[\mu_1(\lambda_{\rm S1}')]\times\mathrm{supp}[\mu_1(\lambda_{\rm S2}')]\times\mathrm{supp}[\pi_{\rm s}(\lambda_{\rm A}')]$ and $\mathrm{supp}[\chi_1(\lambda_{\rm S1}',\lambda_{\rm S2}')]\times\mathrm{supp}[\pi_{\rm f}(\lambda_{\rm A}')]$ in the second equality, and the integration $\int_{\mathrm{supp}[\chi_1(\lambda_{\rm S1}',\lambda_{\rm S2}')]}d\lambda_{\rm S1}'\lambda_{\rm S2}'\chi_2(\lambda_{\rm S1}',\lambda_{\rm S2}')$ is less than or equal to one in the last inequality.

    \subsection*{D.2. Deriving maximum success probability of Eq.~(\ref{gen_pr_cl})}
    From the above calculation, we finally obtain $c\le c^2\alpha_k+1-\alpha_k$ by denoting $c=\Big(\mu_1(\lambda_{\rm S}),\mu_2(\lambda_{\rm S})\Big)_{\Omega_{\rm S}}$. This yields the condition of $\alpha_k$ as
    \begin{eqnarray}\label{ccc}
        \alpha_k\le\frac{1-c}{1-c^2}=\frac{1}{1+c}.
    \end{eqnarray}
    On the top of that, we have to consider a transformation associated with a success outcome
    \begin{eqnarray}
        \mathcal{L}_{\rm s}(\lambda_{\rm S1}',\lambda_{\rm S2}',\lambda_{\rm A}',\lambda_{\rm S})=\alpha_1\mu_1(\lambda_{\rm S1}')\mu_1(\lambda_{\rm S2}')\bar{\mu}_1(\lambda_{\rm S})
        +\alpha_2\mu_2(\lambda_{\rm S1}')\mu_2(\lambda_{\rm S2}')\bar{\mu}_2(\lambda_{\rm S}),
    \end{eqnarray}
    corresponding response function is derived as
    \begin{eqnarray}
        \xi_{\rm s}(\lambda_{\rm S})=\int_{\Omega_{\rm S1}'\times\Omega_{\rm S2}'\times\Omega_{\rm A}'}d\lambda_{\rm S1}'d\lambda_{\rm S2}'d\lambda_{\rm A}'\mathcal{L}_s(\lambda_{\rm S1}',\lambda_{\rm S2}',\lambda_{\rm A}',\lambda_{\rm S})=\alpha_1\bar{\mu}_1(\lambda_{\rm S})+\alpha_2\bar{\mu}_2(\lambda_{\rm S}).
    \end{eqnarray}
    Note that the response function $\xi_{\rm f}(\lambda_{\rm S})=1-\xi_{\rm s}(\lambda_{\rm S})$ also should be non-negative, leading to $\alpha_1+\alpha_2\le\frac{1}{1+c}$. Therefore, the maximum success probability is evaluated as
    \begin{eqnarray}
        \max_{\substack{0\le\alpha_1,\alpha_2\le1 \\ \alpha_1+\alpha_2\le\frac{1}{1+c}}}P_{\rm succ}^{\rm (NC)}=\max_{\substack{0\le\alpha_1,\alpha_2\le1 \\ \alpha_1+\alpha_2\le\frac{1}{1+c}}}{q_1\alpha_1+q_2\alpha_2=\max\{q_1,q_2\}\frac{1}{1+c}}.
    \end{eqnarray}
    This verification is generalized to a task that makes $m\ge n$ copies from $n$ copies. In this case, the condition of Eq.~(\ref{ccc}) is generalized to 
    \begin{eqnarray}
        \alpha_k\le\frac{1-c^n}{1-c^m},
    \end{eqnarray}
    eventually leading to the maximum success probability of  Eq.~(\ref{gen_pr_cl}).

    \subsection*{D.3. Maximum success probability of probabilistic quantum cloning for unequal priors}
    We derive the maximum success probability of the probabilistic quantum cloning in detail~\cite{l.m.duan_cl}, described by the isometric transformation
    \begin{eqnarray}
        \hat{V}|\psi_j\rangle^{\otimes n}=\sqrt{\alpha_j}|\psi_j\rangle^{\otimes m}\otimes|\pi_{\rm s}\rangle+\sqrt{1-\alpha_j}|\chi_j\rangle\otimes|\pi_{\rm f}\rangle,
    \end{eqnarray}
    where $|\pi_{\rm s}\rangle$ and $|\pi_{\rm f}\rangle$ denote orthonormal basis with respect to the success and the failure, respectively. It is straightforward that the above equality leads to
    \begin{eqnarray}
        \langle\psi_j|\psi_k\rangle^{n}-\sqrt{\alpha_j\alpha_k}\langle\psi_j|\psi_k\rangle^m=\sqrt{(1-\alpha_j)(1-\alpha_k)}\langle\chi_j|\chi_k\rangle.
    \end{eqnarray}
    This implies that a Hermitian matrix $[\langle\psi_j|\psi_k\rangle^{n}-\sqrt{\alpha_j\alpha_k}\langle\psi_j|\psi_k\rangle^m]_{j,k=1}^{2}$ is positive-semidefinite~\cite{r.bhatia}. For simplicity, let us assume $\langle\psi_1|\psi_2\rangle=s\in\mathbb{R}$. Then, the positive-semidefiniteness of the Hermitian matrix guides us to that $\alpha_1$ and $\alpha_2$ satisfy
    \begin{eqnarray}
        (1-\alpha_1)(1-\alpha_2)\ge\{s^n-\sqrt{\alpha_1\alpha_2}s^m\}^2.
    \end{eqnarray}
    Intuitively, if $q_1$ ($q_2$) is much larger than $q_2$ ($q_1$), then the optimal solution tends to fulfill $\alpha_2=0$ ($\alpha_1=0$), and the maximum value of $\alpha_1$ ($\alpha_2$) is $1-c^n$ where $c=s^2$~\cite{d.schmid}. Consequently, the maximum success probability is derived as what is written in Eq.~(\ref{cl_qm}).

\section*{Appendix E: Theoretical and experimental discussions \\ of maximal-confidence discrimination}

\subsection*{E.1. Dual problem for evaluating maximal confidence}

We begin by formulating the primal optimization problem for maximal confidence as
\begin{align}
    \text{maximize} \quad & \int_{\Omega} d\lambda \, \widetilde{\mu}_k(\lambda) Q_k(\lambda) \nonumber\\
    \text{subject to} \quad 
    & Q_k(\lambda) \ge 0, \ \forall \lambda \in \Omega, \nonumber\\
    & \int_{\Omega} d\lambda \, Q_k(\lambda) = 1.
\end{align}    
    
To derive the dual problem, we consider the Lagrangian
    \begin{align}
        g(l_k,Z_k(\lambda))&=\max_{Q_k(\lambda)}\left\{\int_{\Omega}d\lambda\widetilde{\mu}_k(\lambda)Q_k(\lambda)+l_k\left(1-\int_{\Omega}d\lambda Q_k(\lambda)\right)+\int_{\Omega}d\lambda Z_k(\lambda)Q_k(\lambda)\right\}\nonumber\\
        &=l_k+\max_{Q_k(\lambda)}\int_{\Omega}d\lambda\left\{\widetilde{\mu}_k(\lambda)-l_k+Z_k(\lambda)\right\}Q_k(\lambda).
    \end{align}
where $Z_k(\lambda) \ge 0$ is a Lagrange multiplier enforcing the positivity constraint.
    
If $\widetilde{\mu}_k(\lambda) - l_k + Z_k(\lambda) \neq 0$ for some $\lambda$, the maximization over $Q_k(\lambda)$ diverges. Therefore, to ensure finiteness, we require
\begin{align}
    \widetilde{\mu}_k(\lambda) - l_k + Z_k(\lambda) = 0,
\end{align}
which implies
\begin{align}
    l_k - \widetilde{\mu}_k(\lambda) \ge 0.
\end{align}

Substituting $\widetilde{\mu}_k(\lambda) = q_k \mu_k(\lambda) / \mu(\lambda)$, we obtain the dual problem
\begin{align}
    \text{minimize} \quad & l_k, \nonumber\\
    \text{subject to} \quad 
    & l_k \mu(\lambda) - q_k \mu_k(\lambda) \ge 0, \ \forall \lambda \in \Omega.
\end{align}

Strong duality holds when the complementary slackness condition is satisfied~\cite{s.boyd,j.bae,h.lee_mc}:
\begin{align}
    \int_{\Omega} d\lambda \, Z_k(\lambda) Q_k(\lambda) = 0.
\end{align}

This condition is equivalent to
\begin{align}
    \int_{\Omega} d\lambda \, \sigma_k(\lambda) \xi_k(\lambda) = 0,
\end{align}
which corresponds to unambiguous discrimination between complementary states $\sigma_k(\lambda)$. Since $k \in \{1,2\}$, this establishes the connection to two-state unambiguous discrimination.

    \begin{figure*}[t]
    \centerline{\includegraphics[width=0.8\linewidth]{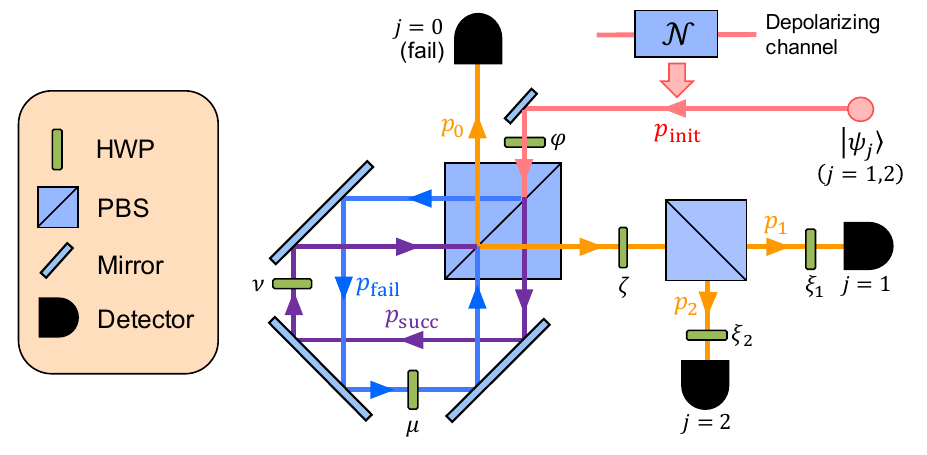}}
    \caption{Proposed experimental setup for unambiguous (maximal-confidence) discrimination between two arbitrary polarized single-photon states in the absence (presence) of a {depolarizing} channel~\cite{y.-c.jeong}. The HWP($\varphi$) is configured according to the prior probabilities of the two states. A displaced Sagnac interferometer, consisting of a PBS, HWP($\mu$), and HWP($\nu$), directs the photon to path $p_0$ with a certain probability, corresponding to a failure outcome. Otherwise, the photon is routed to path $p_1$ or $p_2$, corresponding to outcomes $j=1$ and $j=2$, respectively (PBS: polarizing beam splitter; HWP: half-wave plate).}
    \centering
    \label{figure5}
    \end{figure*}

    \subsection*{E.2. Contextuality enhancement in experimental maximal-confidence discrimination}
    \subsubsection*{1. Concise setup discussion}

We further discuss contextuality enhancement in unambiguous and maximal-confidence discrimination using a linear optical setup, as illustrated in Fig.~\ref{figure5}. In the proposed setup, a sender prepares a single-photon state of the form $|\psi_j\rangle = a_j|H\rangle + b_j|V\rangle$ ($j \in \{1,2\}$) with prior probability $q_j$, under ideal conditions. The receiver employs a displaced Sagnac interferometer to discriminate the prepared states~\cite{m.a.solis-prosser}. After propagation through the interferometer, the joint state of the polarization and path degrees of freedom becomes
\begin{align}
    |\psi_j\rangle \otimes |p_{\rm init}\rangle 
    \rightarrow 
    \sqrt{1-\alpha_j}\,|\phi_j\rangle \otimes |p_{\rm fail}\rangle
    +
    \sqrt{\alpha_j}\,|\chi_j\rangle \otimes |p_{\rm succ}\rangle.
\end{align}

Here, $|p_{\rm fail}\rangle$ and $|p_{\rm succ}\rangle$ denote path states corresponding to failure and success outcomes, respectively. We note that $|\chi_1\rangle$ and $|\chi_2\rangle$ are orthogonal, implying that these polarization states can be perfectly discriminated into path modes labeled by $|p_1\rangle$ and $|p_2\rangle$ using a half-wave plate (HWP) followed by a polarizing beam splitter (PBS). Subsequently, additional HWPs are placed in modes $|p_1\rangle$ and $|p_2\rangle$ to complete the transformation:
\begin{align}
    |\psi_j\rangle \otimes |p_{\rm init}\rangle 
    \rightarrow 
    |\phi_j\rangle \otimes \left( \sqrt{1-\alpha_j}\,|p_0\rangle + \sqrt{\alpha_j}\,|p_j\rangle \right),
    \quad j \in \{1,2\}.
    \label{des_form}
\end{align}

We further introduce an additional HWP with polarization angle $\varphi$, which depends on the prior probabilities. When $\varphi = 0$, the setup reduces to that proposed in Ref.~\cite{m.a.solis-prosser}, which performs optimal unambiguous discrimination between equiprobable pure states. Furthermore, maximal-confidence discrimination~\cite{s.croke,h.lee} can be implemented by incorporating a depolarizing channel~\cite{y.-c.jeong}.

\subsubsection*{2. Unambiguous state discrimination}

We discuss how to implement unambiguous discrimination between two unequiprobable qubit states. We begin by parameterizing the pure states as
\begin{align}
    |\psi_j\rangle = \sqrt{\frac{1+s}{2}}|H\rangle + (-1)^{j+1}\sqrt{\frac{1-s}{2}}|V\rangle,
\end{align}
with $s \in [0,1]$ denoting the overlap between $|\psi_1\rangle$ and $|\psi_2\rangle$.

The interferometer implements the following transformation:
\begin{align}
    |\psi_j\rangle \otimes |p_0\rangle
    &\rightarrow \left\{ h_j(s,\varphi)\cos2\mu|H\rangle - v_j(s,\varphi)\cos2\nu|V\rangle \right\} \otimes |p_{\rm succ}\rangle \nonumber\\
    &\quad + \left\{ v_j(s,\varphi)\sin2\nu|H\rangle + h_j(s,\varphi)\sin2\mu|V\rangle \right\} \otimes |p_{\rm fail}\rangle,
    \label{interf.}
\end{align}
where
\begin{align}
    h_j(s,\varphi) &= \sqrt{\frac{1+s}{2}}\cos2\varphi + (-1)^{j+1}\sqrt{\frac{1-s}{2}}\sin2\varphi, \nonumber\\
    v_j(s,\varphi) &= \sqrt{\frac{1+s}{2}}\sin2\varphi - (-1)^{j+1}\sqrt{\frac{1-s}{2}}\cos2\varphi.
    \label{orth}
\end{align}

The success components must be mutually orthogonal, which leads to
\begin{align}
    h_1(s,\varphi)h_2(s,\varphi)\cos^2 2\mu + v_1(s,\varphi)v_2(s,\varphi)\cos^2 2\nu = 0.
\end{align}

Under this condition, the transformation can be rewritten as
\begin{align}
    |\psi_j\rangle \otimes |p_0\rangle \rightarrow
    \begin{cases}
        \sqrt{\alpha_1}\,|H\rangle \otimes |p_1\rangle
        + \sqrt{1-\alpha_1}\,|\phi_1\rangle \otimes |p_0\rangle, & j=1,\\
        \sqrt{\alpha_2}\,|V\rangle \otimes |p_2\rangle
        + \sqrt{1-\alpha_2}\,|\phi_2\rangle \otimes |p_0\rangle, & j=2,
    \end{cases}
    \label{disc.}
\end{align}
where
\begin{align}
    \alpha_j = h_j^2(s,\varphi)\cos^2 2\mu + v_j^2(s,\varphi)\cos^2 2\nu,
\end{align}
and the post-measurement states are given by
\begin{align}
    |\phi_j\rangle = \frac{v_j(s,\varphi)\sin2\nu|H\rangle + h_j(s,\varphi)\sin2\mu|V\rangle}{\sqrt{v_j^2(s,\varphi)\sin^2 2\nu + h_j^2(s,\varphi)\sin^2 2\mu}}.
    \label{post}
\end{align}

Finally, additional HWPs with angles
\begin{align}
    \xi_1 = \frac{1}{2}\arctan\frac{h_1(s,\varphi)\sin2\mu}{v_1(s,\varphi)\sin2\nu}, 
    \quad
    \xi_2 = \frac{1}{2}\arctan\frac{h_2(s,\varphi)\sin2\nu}{v_2(s,\varphi)\sin2\mu}
\end{align}
complete the transformation in Eq.~(\ref{des_form}).

From Eqs.~(\ref{orth})--(\ref{post}), the optimization problem for the success probability becomes
\begin{align}
    \text{maximize} \quad 
    & P_{\rm succ}(\varphi,\mu,\nu) \nonumber\\
    &= q_1 \alpha_1 + q_2 \alpha_2, \nonumber\\
    \text{subject to} \quad
    & h_1 h_2 \cos^2 2\mu + v_1 v_2 \cos^2 2\nu = 0, \nonumber\\
    & \langle \phi_1 | \phi_2 \rangle = 1.
\end{align}

Since $\alpha_j$ matches the form in Eq.~(\ref{des_form}), the optimal solution attains the IDP bound~\cite{i.d.ivanovic,d.dieks,a.peres,g.jaeger}.

\subsubsection*{3. Maximal-confidence discrimination}

When prepared states are subject to noise, they cannot be distinguished with zero error probability. For instance, consider noisy polarized single-photon states of the form
\begin{align}
    |\psi_j\rangle = \cos\frac{\theta}{2}|H\rangle - (-1)^{j+1}\sin\frac{\theta}{2}|V\rangle,
\end{align}
which can be generated using a depolarizing channel~\cite{y.-c.jeong}. The corresponding mixed states are described by
\begin{align}\label{mix}
    \hat{\rho}_j
    &= p|\psi_j\rangle\langle\psi_j| + (1-p)\frac{\mathbb{I}}{2} = \frac{1+p}{2}|\psi_j\rangle\langle\psi_j|
    + \frac{1-p}{2}|\psi_j^{\bot}\rangle\langle\psi_j^{\bot}|,
\end{align}
where $|\psi_j^{\bot}\rangle$ is orthogonal to $|\psi_j\rangle$.

The key idea is that maximal-confidence discrimination between the mixed states in Eq.~(\ref{mix}) is equivalent to unambiguous discrimination between the complementary states~\cite{h.lee_mc2}
\begin{align}
    |\psi_j^{\rm (c)}\rangle
    = \sqrt{\frac{1+p\cos\theta}{2}}|H\rangle
    - (-1)^{j+1}\sqrt{\frac{1-p\cos\theta}{2}}|V\rangle.
\end{align}

This suggests that maximal-confidence discrimination can be implemented by configuring the setup in Sec.~E.2.1 to unambiguously discriminate these complementary states.

Within the interferometer, both $|\psi_j\rangle$ and $|\psi_j^{\bot}\rangle$ evolve as
\begin{align}
    |\psi_j\rangle \rightarrow |\Psi_j\rangle
    &= \left\{ h_j(\theta,\varphi)\cos2\mu|H\rangle - v_j(\theta,\varphi)\cos2\nu|V\rangle \right\} \otimes |p_{\rm succ}\rangle \nonumber\\
    &\quad + \left\{ v_j(\theta,\varphi)\sin2\nu|H\rangle + h_j(\theta,\varphi)\sin2\mu|V\rangle \right\} \otimes |p_{\rm fail}\rangle, \nonumber\\
    |\psi_j^{\bot}\rangle \rightarrow |\Psi_j^{\bot}\rangle
    &= \left\{ \bar{h}_j(\theta,\varphi)\cos2\mu|H\rangle - \bar{v}_j(\theta,\varphi)\cos2\nu|V\rangle \right\} \otimes |p_{\rm succ}\rangle \nonumber\\
    &\quad + \left\{ \bar{v}_j(\theta,\varphi)\sin2\nu|H\rangle + \bar{h}_j(\theta,\varphi)\sin2\mu|V\rangle \right\} \otimes |p_{\rm fail}\rangle,
    \label{tr_mc}
\end{align}
where
\begin{align}
    h_j(\theta,\varphi) &= \cos\frac{\theta}{2}\cos2\varphi - (-1)^{j+1}\sin\frac{\theta}{2}\sin2\varphi, \nonumber\\
    v_j(\theta,\varphi) &= \cos\frac{\theta}{2}\sin2\varphi + (-1)^{j+1}\sin\frac{\theta}{2}\cos2\varphi, \nonumber\\
    \bar{h}_j(\theta,\varphi) &= \sin\frac{\theta}{2}\cos2\varphi + (-1)^{j+1}\cos\frac{\theta}{2}\sin2\varphi, \nonumber\\
    \bar{v}_j(\theta,\varphi) &= \sin\frac{\theta}{2}\sin2\varphi - (-1)^{j+1}\cos\frac{\theta}{2}\cos2\varphi.
\end{align}

Conditioned on detecting the path state $|p_{\rm succ}\rangle$, the polarization is measured in the orthonormal basis
\begin{align}
    |\pi_k^{\rm (c)}\rangle
    = \frac{
    h_k^{\rm (c)}(\theta,\varphi)\cos2\mu|H\rangle
    + v_k^{\rm (c)}(\theta,\varphi)\cos2\nu|V\rangle
    }{
    \sqrt{
    h_k^{\rm (c)2}(\theta,\varphi)\cos^2 2\mu
    + v_k^{\rm (c)2}(\theta,\varphi)\cos^2 2\nu
    }},
    \label{pr_c}
\end{align}
with
\begin{align}
    h_k^{\rm (c)}(\theta,\varphi)
    &= \sqrt{\frac{1+p\cos\theta}{2}}\cos2\varphi
    - (-1)^{k+1}\sqrt{\frac{1-p\cos\theta}{2}}\sin2\varphi, \nonumber\\
    v_k^{\rm (c)}(\theta,\varphi)
    &= \sqrt{\frac{1+p\cos\theta}{2}}\sin2\varphi
    + (-1)^{k+1}\sqrt{\frac{1-p\cos\theta}{2}}\cos2\varphi.
\end{align}

Using Eqs.~(\ref{tr_mc}) and (\ref{pr_c}), the probability of obtaining outcome $k$ from $\hat{\rho}_j$ is given by
\begin{align}
    \mathrm{Pr}[k|\hat{\rho}_j]
    = \frac{
    (1+p)f_{jk}(\theta,\varphi,\mu,\nu)
    + (1-p)\bar{f}_{jk}(\theta,\varphi,\mu,\nu)
    }{
    2\left\{
    h_k^{\rm (c)2}(\theta,\varphi)\cos^2 2\mu
    + v_k^{\rm (c)2}(\theta,\varphi)\cos^2 2\nu
    \right\}
    },
\end{align}
where 
\begin{align}
    f_{jk}(\theta,\varphi,\mu,\nu)=\left\{h_j(\theta,\varphi)h_k^{\rm (c)}(\theta,\varphi)\cos^22\mu+v_j(\theta,\varphi)v_k^{\rm (c)}(\theta,\varphi)\cos^22\nu\right\}^2,\nonumber\\
    \bar{f}_{jk}(\theta,\varphi,\mu,\nu)=\left\{\bar{h}_j(\theta,\varphi)h_k^{\rm (c)}(\theta,\varphi)\cos^22\mu+\bar{v}_j(\theta,\varphi)v_k^{\rm (c)}(\theta,\varphi)\cos^22\nu\right\}^2.
\end{align}

Consequently, the maximal-confidence optimization problem becomes
\begin{align}
    \text{maximize} \quad 
    & C_k(\varphi,\mu,\nu)
    = \frac{q_k\,\mathrm{Pr}[k|\hat{\rho}_k]}{q_1\,\mathrm{Pr}[k|\hat{\rho}_1] + q_2\,\mathrm{Pr}[k|\hat{\rho}_2]}, \nonumber\\
    \text{subject to} \quad
    & h_1^{\rm (c)} h_2^{\rm (c)} \cos^2 2\mu + v_1^{\rm (c)} v_2^{\rm (c)} \cos^2 2\nu = 0, \nonumber\\
    & \langle \phi_1 | \phi_2 \rangle = 1.
\end{align}

For example, for $\theta = 0.42\pi$, $p = 0.58$, and $q_1 = 0.65$, the optimal confidences are $0.870719$ and $0.661335$, respectively, in agreement with the theoretical predictions. This confirms that the proposed setup enables an experimental realization of contextuality-enhanced maximal-confidence discrimination.

\end{widetext}

\end{document}